\documentstyle[12pt, aasms4, graphics]{article}

\begin{document}

\title{GAMMA RAY BURST HOST GALAXIES HAVE `NORMAL' LUMINOSITIES}

\author{Bradley E. Schaefer}
\affil{Yale University, Physics Department, 260 Whitney, New Haven CT 06511}

\baselineskip 24pt

\begin{abstract}
 
The galactic environment of Gamma Ray Bursts can provide good evidence
about the nature of the progenitor system, with two old arguments implying
that the burst host galaxies are significantly subluminous.  New data and
new analysis have now reversed this picture:  (A) Even though the first
two known host galaxies are indeed greatly subluminous, the next eight
hosts have absolute magnitudes typical for a population of field galaxies.
A detailed analysis of the 16 known hosts (ten with red shifts) shows them
to be consistent with a Schechter luminosity function with $R^{*} = -21.8 \pm 1.0$, 
as expected for normal galaxies.  (B) Bright bursts from the Interplanetary
Network are typically 18 times brighter than the faint bursts with red
shifts, however the bright bursts do not have galaxies inside their error
boxes to limits deeper than expected based on the luminosities for the two
samples being identical.  A new solution to this dilemma is that a broad
burst luminosity function along with a burst number density varying as the
star formation rate will require the average luminosity of the bright
sample ($>$$6 \times 10^{58} ph \cdot s^{-1}$ or $>$$1.7 \times 10^{52} \cdot erg 
\cdot s^{-1}$) to be much greater than the average luminosity of the faint 
sample ($\sim 10^{58} ph \cdot s^{-1}$ or $\sim 3 \times 10^{51} erg \cdot 
s^{-1}$).  This places the bright bursts at distances for which host galaxies 
with a normal luminosity will not violate the observed limits.  In conclusion,
all current evidence points to GRB host galaxies being normal in
luminosity.
\end{abstract}


\clearpage
\section{Introduction}

 	The key puzzle of Gamma Ray Bursts (GRBs) is the nature of their
central engine.  Observations of the burst afterglows can provide some
information about the cause of the burst, but this is limited since the
explosion will destroy much evidence.  Another type of knowledge that will
be useful is to identify the environment of the burster, since this will
provide information about the progenitor.  For example, if GRBs appear
outside galaxies then models with binary systems containing collapsed
stars that can have high ejection velocities will be preferred, while if
GRBs appear preferentially in high luminosity galaxies with rapid star
formation then models with very massive progenitors will be preferred.  So
useful questions are ``Do GRBs appear within normal galaxies?'' and ``What is
the luminosity function of the GRB host galaxies?''

	In the past, two arguments have been presented that made strong
cases that most GRBs appeared either outside normal galaxies, in
systematically subluminous hosts, or at high luminosities (Schaefer 1999;
Band, Hartmann, \& Schaefer 1999).  The first argument is that GRB970228
and GRB970508 (the first two identified GRB hosts) are in the bottom $\sim 1\%$
of the luminosity weighted Schechter luminosity function, with this result
being unlikely unless the GRB hosts are systematically subluminous.  The
second argument is that a dozen very bright bursts seen with the
Interplanetary Network (IPN; Hurley 1986; Hurley et al. 1993) have no
galaxies in their small error boxes to B magnitudes from 20 to 24, whereas
the hosts should have been easily visible if the bursters reside in normal
host galaxies for the luminosities allowed by LogN-LogP studies.

	In the past year, new burst red shifts have greatly changed the
situation from that presented in the previous paragraph.  Also, I here
propose an alternative solution for the lack of sufficiently-bright hosts
for the bright bursts.  This paper presents these two new analyses, with
the conclusion that GRBs reside in normal host galaxies whose luminosities
are distributed as a normal Schechter luminosity function.

\section{Hosts of Faint Bursts}

	The first two discovered GRB hosts (for GRB970228 and GRB970508)
are galaxies at the bottom of the luminosity function.  But now we have
data for hosts on sixteen GRBs with optical transients or radio transients
(OT/RT) to provide arc-second positions and ten of these have measured red
shifts (see Table 1).  This much larger sample can answer the question of
``What is the luminosity function for the host galaxies of faint bursts?''

	An approximate answer to this question can be obtained by merely
examining the derived absolute magnitudes of the hosts as taken from Table
1.  We see that the first two GRB hosts are fortuitously the least
luminous hosts by about one magnitude.  This means that the early argument
for subluminous hosts based on GRB970228 and GRB970508 is wrong due to a
rather unlikely coincidence.  Further, we see that the typical R-band
absolute magnitude is around -21, a value which is comparable to the $R^{*}$
value characteristic of the R-band Schechter luminosity function.  $R^{*}$ is
approximately -21.2 mag in the local vicinity for a Hubble constant of $65
km \cdot s^{-1} \cdot Mpc^{-1}$ (Lin et al. 1996).  So to first order, the 
host galaxies of faint GRBs have a normal luminosity.

	However, a variety of effects and biases can affect this
conclusion:  The probability of a detected burst yielding a red shift and
an apparent magnitude for the host depends on the burst distance, the
burst luminosity, and the host's absolute magnitude.  So our sample in
Table 1 will be biased towards luminous hosts for which a red shift is
more likely to be measured.    Also, the bursts in Table 1 have a typical
red shift of $\sim 1$, so that effects due to the values of cosmological
parameters ($\Omega_{m}, \Omega_{\Lambda}$) and the K-corrections for 
both bursts and hosts will affect the conclusion.  Hence, the analysis 
given in the previous paragraph needs to be improved.

	An improved analysis is to model all these effects and biases with
a Monte Carlo calculation to produce a simulated catalog of bursts
containing subsets with red shifts and host apparent magnitudes.  I have
adopted a Hubble constant of $65 km \cdot s^{-1} \cdot Mpc^{-1}$ in a 
flat Universe with $\Omega_{m} = 0.3$.  I take the burst number density to 
follow the star formation rate as given by Madau, Pozzetti, \& Dickinson (1998).
  The burst luminosity function is taken as the usual truncated power law 
with slope -2, dynamic range of 1000, and a minimum luminosity of $10^{57} ph 
\cdot s^{-1}$, to be consistent with the observed time dilation, red shifts, 
and LogN-LogP curve (Deng \& Schaefer 1999) as well as the light curve 
variability (Ramirez-Ruiz \& Fenimore 1999).  The host luminosity function 
was taken to have the shape of the Schechter luminosity function with slope 
$\alpha = -1$ and a characteristic R-band absolute magnitude, $R^{*}$, which 
is a free parameter.  Based on the events in Table 1 and the LogN-LogP curve, I 
will approximate the probability of getting an arc-second position for an 
observed burst as rising linearly from zero for $P_{256}$ values ranging 
between 0.5 to 5.5 $ph \cdot s^{-1} \cdot cm^{-2}$.  Similarly, the probability 
of measuring an apparent magnitude for a host galaxy is taken to be 0.7 if the 
burst has an arc-second position and a host brighter than R=25.7 mag.  The 
K-corrections for the host are taken for those of an Sb galaxy as given by 
Rocca-Volmerange \& Guideroni (1988).  The K-corrections for the burst are 
taken for a count spectrum varying as $E^{-2}$ (Schaefer et al 1994; 1998).

	What parameters should be used to compare the model with the
observations?  A comparison of apparent magnitudes allows for more
measured values from Table 1 than would a use of absolute magnitudes.
Reasonable aggregate parameters for model comparisons are the median and
the standard deviation for the apparent magnitudes of detected bursts
(24.88 and 1.49 mag; see Table 1).

	  Few observed apparent magnitudes are currently known, so the
shape of the host luminosity function cannot yet be well constrained.
Nevertheless, the observed scatter in $R_{host}$ is a function of the shape and
can indicate consistency with the luminosity-weighted Schechter luminosity
function adopted.  For reasonable models, the typical standard deviation
of $R_{host}$ is 2.4 mag, although this varies widely for samples of 16 bursts.
The observed standard deviation (1.49 mag) is not surprisingly smaller
than these model values, so as yet there is no inconsistency with the
shape of the Schechter function.  The $<$$R_{host}$$>$ value for 16 bursts 
varies with a standard deviation of $2.4/\sqrt{16}$ or 0.60 mag, so the target 
for the model is $24.88 \pm 0.60$ mag.

	For what values of $R^{*}$ does the model reproduce the observed
distribution of host apparent magnitudes?  Figure 1 displays the model
predictions as a function of the adopted $R^{*}$.  An acceptable range of $R^{*}$ 
is then from -21.2 to -22.4 mag, with the best value being around -21.8 mag.

	However, uncertainties in the model input parameters will enlarge
the acceptable range of $R^{*}$.  This can be quantified by calculating the
change in the model $<$$R_{host}$$>$ when one input parameter is changed 
over some plausible range (with the luminosity function shifted such that the
observed $<$$LogL$$>$ is held constant).  A change in $\Omega_{m}$ from 0.3 
to 1.0 makes the average $R_{host}$ fainter by 0.34 mag.  A change of the 
Hubble constant
will change both the model $<$$R_{host}$$>$ and the $R^{*}$ value for normal 
galaxies to the same degree, with these effects canceling out.  A change in 
the average slope of the GRB count spectrum from $E^{-2}$ to $E^{-1.5}$ changes 
$<$$R_{host}$$>$ by less than 0.1 mag.  A shift in the intrinsic burst 
luminosity function by a factor of two changes $<$$R_{host}$$>$ by 0.1 mag and 
0.3 mag for brighter and fainter bursts respectively.  Even large changes in 
the shape of the burst luminosity function move $<$$R_{host}$$>$ by less than 
0.2 mag.  So an uncertainty of $\sim 0.4$ mag in the model $<$$R_{host}$$>$ 
arises from uncertainties in the model input parameters.  Then, the range of 
acceptable $R^{*}$ values increases to from -20.8 to -22.8, so the final model 
estimate of $R^{*}$ is $-21.8 \pm 1.0$ mag.

	This derived $R^{*}$ value is easily consistent with normal galaxies,
yet is inconsistent with greatly subluminous hosts.  The uncertainty in R*
is larger than desirable due to the few available GRB hosts known to date
and to significant uncertainties in the conditions of the high red shift
Universe.  These will be improved with time.  For now, the conclusion is
that GRBs appear to have host galaxies of normal luminosity.

\section{Hosts of Bright Bursts}

	The bursts with optical or radio transients are typically rather
faint, with the median $P_{256}$ being only a factor of 3 above the BATSE
completeness threshold.  These GRBs are greatly fainter than the bursts
positioned with the IPN (see Fig. 2).  For a fair comparison, the sixteen
OT/RT bursts (Table 1) can be compared with the sixteen IPN bursts with
the smallest error boxes (Schaefer et al. 1998).  The median IPN burst is
18 times brighter than the median OT/RT burst.

	For many reasonable models, the IPN bursts should thus be $\sim 4$ times
closer than the OT/RT events and then will be substantially immune to many
problems that plague the interpretation of the high red shift OT/RT events
(uncertainties in the K-corrections, the cosmological parameters, the dust
extinction, and the galaxy luminosity function).  For some purposes, the
IPN burst sample might then be more important than the OT/RT sample
because the low red shift Universe can be readily interpreted.

	Schaefer (1999) and Band, Hartmann, \& Schaefer (1999) both examine
the limits on Rhost for the IPN GRBs, with the conclusion that the hosts
can have normal luminosities (i.e., be drawn from the usual
luminosity-weighted Schechter luminosity function) only if the average
burst luminosity is greater than $6 \times 10^{58} ph \cdot s^{-1}$ (LogL=58.8).  
This directly contradicts fits to the LogN-LogP curve (Horv\'{a}th, M\'{e}sz\'{a}ros, 
\& M\'{e}sz\'{a}ros 1996), the time dilation of burst light curves (Deng \& Schaefer 1999), 
as well as the observed luminosities for the OT/RT bursts (see Table 1).  
Possible solutions to this dilemma were that the GRBs were ejected from their 
birth galaxy or that the host galaxies are systematically subluminous for some
reason.  Neither solution now seems plausible.

	I would like to point out another solution which fits well with
currently popular ideas.  The dilemma arises because the bright IPN bursts
were plausibly assumed to have the same mean luminosity as the faint OT/RT
bursts.  However, if GRBs simultaneously have a broad luminosity function
and their number density increases greatly with red shift, then the bright
bursts will have a much greater average luminosity than will faint bursts.
That is, if the OT/RT events have $LogL \approx 58.0$ while the IPN events have
$LogL$$>$$58.8$, then the host galaxies of the IPN bursts will have $R_{host}
\sim 24$ and be fully consistent with the limits in Schaefer et al. (1998). 

	To provide a quantitative evaluation of this idea, I have
calculated the average luminosities and red shifts for bursts with peak
fluxes brighter than some threshold for a variety of burst luminosity
functions.  The required integrals were performed numerically for red
shifts over the range 0-6 with bins of 0.01.  The luminosity distances
were calculated for a flat Universe with $\Omega_{m} = 0.3$ (hence 
$\Omega_{\Lambda} = 0.7$) with a Hubble constant of $65 km \cdot s^{-1} \cdot 
Mpc^{-1}$.  The star formation rate was taken
from Madau, Pozzetti, \& Dickinson (1998) as deduced from the rest-frame UV
luminosity density.  The K-corrections were made assuming that the average
burst count spectrum is an $E^{-2}$ power law, as indicated in Fig. 46 of
Schaefer et al. (1994) and Figure 4 of Schaefer et al. (1998).  The burst
luminosity function was taken either as a log-normal distribution or as a
truncated power law.  The characteristic widths of these were allowed to
vary widely, but the average luminosity was set such that $<$$LogL$$>$ for a
population observed with $P_{256}$$>$$1 ph \cdot s^{-1}$ was 58.34 (see Table 1).  
Figure 3 displays the results for the truncated power law from the previous 
section as well as for two widths of a log-normal luminosity function.  

	Both power law and log-normal distributions give similar results,
in that samples of bright bursts will be much more luminous than samples
of dim bursts.  The one-sigma scatter in the observed LogL values varies
from 0.5 to 0.9, which is comparable to that seen in Table 1.  The mean
red shift of bright burst samples is much higher than would be expected
from simple scaling by $P_{256}^{-0.5}$ from the red shift of a faint burst
sample, for example the log-normal luminosity function with width 1.0 has
a ratio of $<$$z$$>$ equal to 2.0 for samples with $P_{256}$ greater than 1.0 
and $30 ph \cdot s^{-1} \cdot cm^{-2}$.  In the extreme case of a very broad 
power law with $\sim L^{-2}$, the $<$$z$$>$ will be roughly a constant.

	An interesting result from these calculations is that the
luminosity function of the {\it observed} GRBs is roughly log-normal for any
broad {\it intrinsic} shape.  That is, both log-normal and power law input
functions produce apparently log-normal output functions.  With a broad
log-normal input function, the observed $<$$LogL$$>$ will be over ten times
larger than the intrinsic $<$$LogL$$>$ so that the shape of the intrinsic
distribution near and below its peak is irrelevant.  This result is due to
a cutoff in the observed events on the low luminosity side by the small
volume of space near enough for weak bursts to be detectable and a cutoff
on the high luminosity side by the rapid decrease in the number of strong
bursts.  Unfortunately, an implication is that an approximately log-normal
observed luminosity distribution (see Table 1 or Ruiz-Ramirez \& Fenimore
1999) can only tell us that the intrinsic luminosity function is broad.

	For broad luminosity functions, the $<$$LogL$$>$ for observed bursts is
determined by the overall slope of the intrinsic luminosity function.
Thus for $LogL \sim 58$, the GRB luminosity function must scale close to $
L^{-2}$ regardless of the behavior at high and low luminosity.  

	The primary point of Fig. 3 is that the average luminosity of the
bright bursts is greatly larger than for the faint bursts.  The most
important comparison is for bursts with $P_{256}$$>$$0.85 ph \cdot s^{-1}
\cdot cm^{-2}$ (the BATSE completeness threshold which is relevant for 
logN-LogP studies and for the OT/RT bursts in Table 1) versus bursts with 
$P_{256}$$>$$30 ph \cdot s^{-1} \cdot cm^{-2}$ (for the IPN bursts).  For 
the three broad luminosity functions in Fig. 3, the ratio of luminosities 
for these two thresholds is 8.3, 4.0, and 13.5.  That is, the average 
luminosity of the IPN bursts is roughly an order of magnitude brighter than 
for the OT/RT and BATSE bursts.  This means that for OT/RT and BATSE bursts 
with $<$$LogL$$>\sim 58.0$ (Horv\'{a}th, M\'{e}sz\'{a}ros, \& M\'{e}sz\'{a}ros 1996; Deng 
\& Schaefer 1999; Ruiz-Ramirez \& Fenimore 1999; Table 1) then
the IPN bursts likely have $<$$LogL$$> \sim 59.0$.  This is completely 
consistent with the lack of hosts in IPN boxes to deep limits (Schaefer 1999).

	In summary, the two original arguments that hosts are subluminous
are now shown to be incorrect, with the new conclusion that hosts are just
normal galaxies without need of any special environment for the GRB
progenitors.

\clearpage

\begin{table}
\begin{center}
\caption{Host galaxy data for bursts with optical or radio positions.}
\begin{tabular}{llllllll}
\hline
GRB & $P_{256}^{a}$ & Transient$^{b}$ & $z(Method)^{c}$ & $Log_{10}L^{d}$ 
& $R_{host}^{e}$ & $M_{host}^{f}$ & $Ref.$ \\
      & $ph \cdot s^{-1} \cdot cm^{-2}$ & & & $ph \cdot s^{-1}$ & mag & mag & \\		
\hline
970228 & 10.0 & O & 0.695 (He) & 58.38 & 24.96 & -18.92 & g \\
970508 & 1.2  & O, R & 0.835 (Aeax) & 57.65 & 25.65 & -19.13 & h \\
970828 & $\sim 1.5$ & R	& 0.958 (He) & 57.90 & 24.41 & -20.93 & i \\
971214 & 2.3 & O & 3.418 (He) & 59.45 & 25.56 & -24.46 & j \\
980326 & $\sim 3$ & O & $\cdots$ & $\cdots$ & $>$$27.1$ & $\cdots$ & k \\
980329 & 13.3 & O, R & $\cdots$ & $\cdots$ & 25.94 & $\cdots$ & l \\
980425 & 1.1 & SN & 0.0085 (Ha) & 53.24	& $\sim 13$ & $\sim -20$ & m \\
980519 & 4.7 & O, R & & $cdots$ & 25.40	& $\cdots$ & n \\
980613 & 0.5 & O & 1.097 (He) & 57.57 & 23.85 & -22.07 & o \\
980703 & 2.6 & O, R & 0.966 (Aea) & 58.04 & 22.43 & -22.94 & p \\
981226 & $\sim 0.4$ & R & & $\cdots$ & 24.79 & $\cdots$	& q \\
990123 & 16.6 & O, R & 1.600 (Aa) & 59.50 & 23.73 & -23.85 & r \\
990308 & 1.6 & O &  & $\cdots$ & $>$$25.7$ & $\cdots$ & s \\	
990506 & 22.2 & R &  & $\cdots$ & 24.7 & $\cdots$ & t \\
990510 & 10.2 & O, R & 1.619 (Aa) & 59.30 & $>$$27.0$ & $>$$-20.59$ & u \\
990712 & $\sim 2.3$ & O & 0.430 (Aea) & 57.25 & 21.71 & -20.63 & v \\
Median: & 2.5 & & 0.96 & 57.97 & 24.88 & -20.78$^{y}$ & \\	
Average: & 3.1$^{w}$ & & 1.09$^{wx}$ & 58.34$^{x}$ & 24.86$^{x}$ & -21.35$^{y}$ & \\
\hline
\end{tabular}
\end{center}
\end{table}
\clearpage
Notes: $^{a}$Peak flux in the 50-300 keV band over a 256 ms time interval.
Values are from the BATSE online catalog (http://gammaray.msfc.nasa.gov/batse/grb/) 
or by scaling from the SAX peak flux.  $^{b}$O indicates that an optical (or 
near infrared) transient was seen, R is for a radio transient, and SN is for 
a supernova.  $^{c}$A or H indicates whether the afterglow or the host galaxy 
was used to measure the red shift; while a, e, or x indicates optical 
absorption lines, optical emission lines, or x-ray emission lines.  $^{d}$The 
burst luminosity is calculated from $P_{256}\cdot 4\pi D_{l}^{2}$, where $D_{l}$ is the luminosity 
distance for a $\Omega_{m} = 0.3$ flat Universe (hence $\Omega_{\Lambda} = 0.7$) and 
a Hubble constant of $65 km \cdot s^{-1} \cdot Mpc^{-1}$.  To find an average 
equivalent luminosity with units $erg \cdot s^{-1}$ from 30-2000 keV (Fenimore 
et al. 1993), divide by approximately $3.6 \times 10^{6}$ (or subtract 6.55 
from the logarithm).  $^{e}$The R-band magnitude for the host galaxy after a 
correction for the absorption from our Milky Way galaxy.  $^{f}$The absolute 
R-band magnitude of the host galaxy based on the tabulated magnitudes and red 
shifts.  K-corrections were applied for Sb galaxies with no E-corrections as 
taken from Rocca-Volmerange \& Guideroni (1988).  For GRB971214 at z=3.412, I 
adopt a K-correction of 2.5 mag.  At z=1, the range of K-corrections is 0.7 
mag over the classes of spiral galaxies.  $^{g}$Djorgovski et al. 1999b; 
Fruchter et al. 1999b; Castander \& Lamb 1999.  $^{h}$Metzger et al. 1997; 
Bloom et al. 1998b.  $^{i}$ Djorgovski 1999.  $^{j}$Kulkarni et
al. 1998.  $^{k}$ Bloom et al. 1999b.  $^{l}$ Djorgovski 1999.  $^{m}$ Galama 
et al. 1998.  $^{n}$ Bloom et al. 1998a.  $^{o}$ Djorgovski et al. 1999a.  
$^{p}$ Djorgovski et al. 1998; Vreeswijk et al. 1999b.  $^{q}$ Frail et al. 
1999.  $^{r}$ Hjorth et al. 1999; Halpern et al. 1999.  $^{s}$ Schaefer et al. 
1999.  $^{t}$ Bloom et al. 1999a.  $^{u}$ Vreeswijk et al. 1999a; Fruchter et 
al. 1999a.  $^{v}$ Galama et al. 1999; Kemp et al. 1999.  $^{w}$ The geometric 
mean was used.  $^{x}$ The averages exclude GRB980425 since its red shift and 
luminosity can plausibly be considered to be from a separate population.  
$^{y}$ With E-corrections for the host galaxy from Rocca-Volmerange
\& Guideroni (1988), the median $M_{host}$ is -20.06, the average Mhost is
-19.90 and the standard deviation of Mhost is 1.42 mag.

\clearpage

\begin{figure}
\begin{center}
\resizebox{12.5cm}{10cm}{\includegraphics{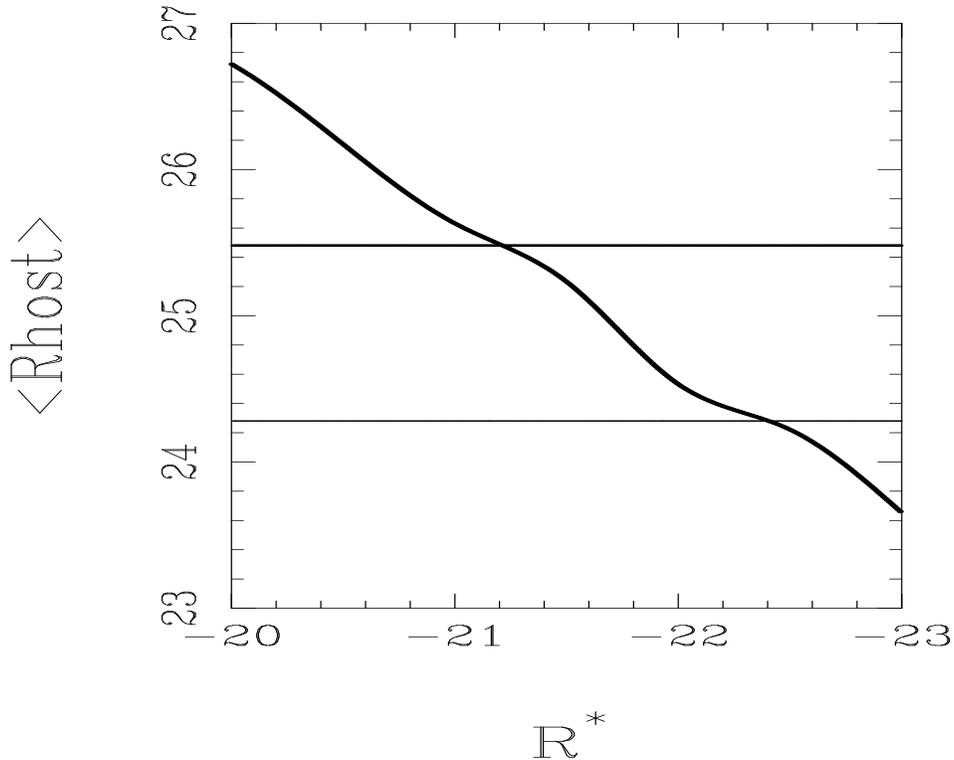}}
\caption{Average host magnitude for both observed bursts and the 
model.  The analysis of host magnitudes has moved past consideration of just 
the first two known hosts as well as simple calculations of the average host
absolute magnitude.  The graph below compares the results from realistic
Monte Carlo simulated burst catalogs (the sloped curve) with the range of
uncertainty for the observed values (between the two horizontal lines;
$24.88 \pm 0.60$ mag).  The one free parameter in the model is the $R^{*}$ 
value which characterizes the host galaxy luminosity, with $R^{*} = -21.2$ 
(for a Hubble constant of $65 km \cdot s^{-1} \cdot Mpc^{-1}$).  The 
acceptable range for is then $-21.2$$>$$R^{*}$$>$$-22.4$ for the adopted model 
parameters, although this range is increased to $-20.8$$>$$R^{*}$$>$$-22.8$ 
when allowance is made for plausible uncertainties in the adopted model 
parameters.  From this, we see that GRB hosts are apparently of normal 
luminosity and certainly not greatly subluminous on average (despite the 
first two known hosts being greatly subluminous).}
\end{center}
\end{figure} 

\clearpage

\begin{figure}
\begin{center}
\resizebox{12.5cm}{12.5cm}{\includegraphics{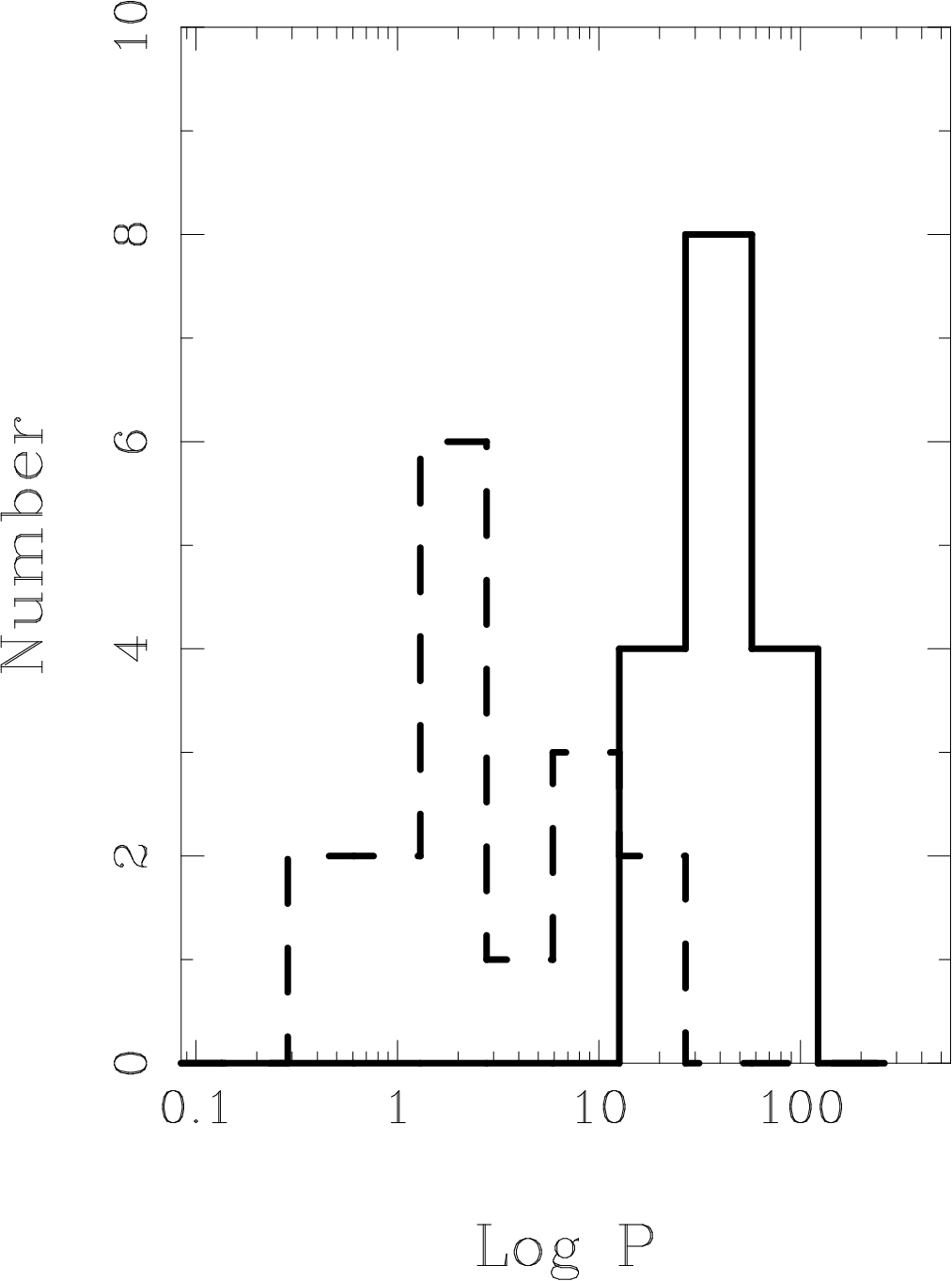}}
\caption{The IPN bursts are $\sim 18$ times brighter than the GRB/OT bursts.  
This histogram shows the distribution of $P_{256}$ for two samples 
of sixteen bursts; (a) the smallest IPN bursts shown with the solid line and 
(b) the OT/RT bursts with arc-second positions from Table 1 shown with the 
dashed line.  The two distributions are nearly separated, with the OT/RT 
bursts greatly fainter than the IPN bursts.  The medians of the two 
distributions have a ratio of 18.}
\end{center}
\end{figure} 

\clearpage

\begin{figure}
\begin{center}
\resizebox{12.5cm}{10cm}{\includegraphics{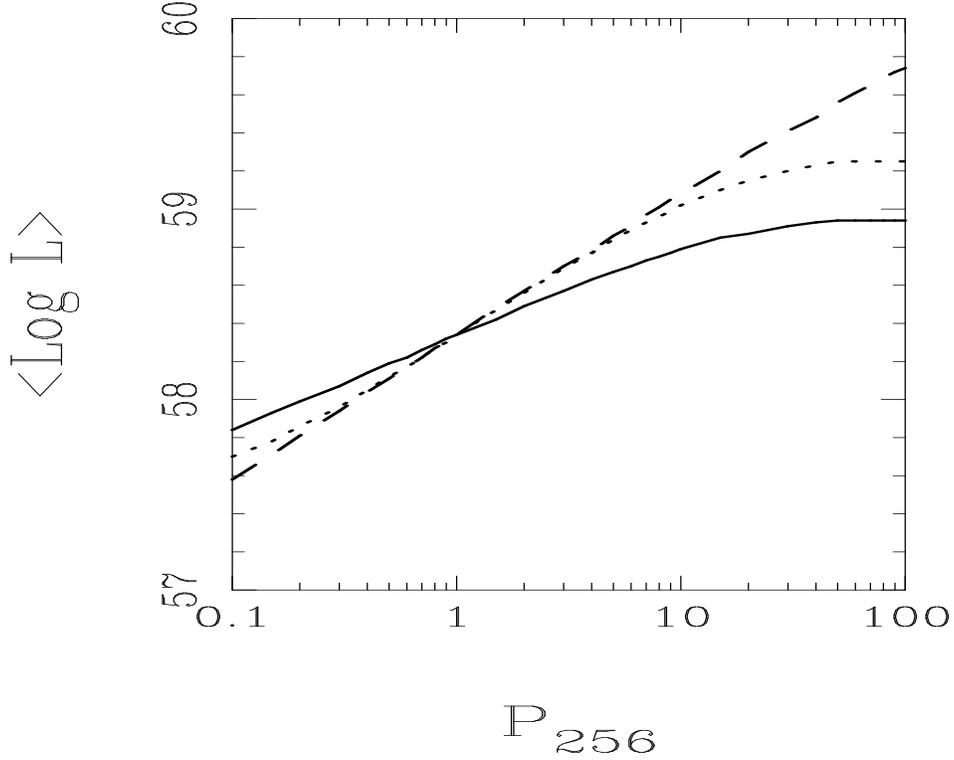}}
\caption{Bright bursts are much more luminous than faint bursts.
The average luminosity ($<$$log_{10}(L)$$>$ with units of $ph \cdot s^{-1}$) 
for observed bursts depends sharply on the threshold $P_{256}$ (in $ph \cdot
s^{-1} \cdot cm^{-2}$ from 50-300 keV) value for the sample despite a constant 
intrinsic luminosity function.  The intermediate curve is for a truncated 
power law luminosity function as used in section 2.  Two curves are for a 
log-normal intrinsic luminosity function of width 1.0 (i.e. the typical 
dispersion is a factor of ten; the shallow curve), and of width 2.0 (i.e. 
the typical dispersion is a factor of one hundred; the steep curve).  A 
comparison between BASTE or OT/RT bursts (threshold $P_{256} = 0.85 ph \cdot
 s^{-1} \cdot cm^{-2}$) and IPN bursts (threshold $P_{256} = 30 ph \cdot s^{-1}
\cdot cm^{-2}$) must account for the factor of $\sim 10$ difference in average
luminosity.  This realization resolves the discrepancy that faint bursts
have $<$$LogL$$>$ around 58.0 (based on time dilation of the light curves, the
LogN-LogP curve, and the few known red shifts) while bright bursts have
$<$$LogL$$>$ greater than 58.8 (based on the lack of host galaxies to deep
limits).}
\end{center}
\end{figure}

\end{document}